# PREPARED OR UNPREPARED? EVALUATING HEALTHCARE WORKFORCE READINESS FOR CLINICAL ADOPTION OF ARTIFICIAL INTELLIGENCE IN NIGERIA

Abbas M. Rabiu[1] [0000-0002-7139-9341] Abdulrazaq A. Zubair[2,3] [0009-0007-8010-3999] Ummulkhairi Ibrahim[1,3] [0009-0005-9542-8011] Tolulope Olusuyi[4] [0009-0007-7182-0785] Shaheeda Farouq[5] [0000-0001-7428-2985] Safwan M. Dafi[5] [0009-0005-4518-2770] Adaobi C. Emegoakor[6] [0009-0008-2874-6708] Yewande Gbadamosi[7] [0009-0002-7058-4905] Maruf Adewole[4] [000-0002-6562-2834]

[1] College of Health Sciences, Bayero University Kano, Nigeria
[2] Department of Radiography and Radiation Sciences, Federal University of Health Sciences, Azare, (FUHSA) Nigeria
abdulrazaq.zubair@fuhsa.edu.ng
[3] African Institute for Research Advancement & Innovation (AIRA AFRICA)
[4] Medical Artificial Intelligence Laboratory (MAI Lab), Lagos, Nigeria
[5] Department of Radiology, Aminu Kano Teaching Hospital, Kano, Nigeria
[6] Department of Radiology, Nnamdi Azikiwe University Teaching Hospital, Nnewi, Nigeria
[7] Department of Radiology, Lagos State Teaching Hospital, Lagos, Nigeria

**Abstract.** Artificial intelligence (AI) is increasingly integrated into healthcare systems worldwide, yet its successful clinical adoption depends critically on workforce readiness, particularly in low- and middle-income countries (LMICs) where infrastructural and training gaps persist. This cross-sectional study evaluated awareness, attitudes, preparedness, and barriers to AI adoption among 761 healthcare professionals across multiple disciplines and practice settings in Nigeria. Data were collected between December 2025 and March 2026 using a structured, validated questionnaire. Overall awareness of AI in healthcare was high (92.6%); however, objective knowledge and self-reported preparedness remained limited, with 40.9% reporting low or very low knowledge and only 63.0% feeling adequately prepared. Willingness to adopt AI was high: 92.5% expressed interest in training, and 78.7% supported inclusion of AI education in undergraduate curricula. Key barriers included lack of training (84.7%), poor infrastructure (71.1%), high cost of AI tools (61.0%), fear of job displacement (60.6%), ethical concerns (52.9%), and data privacy concerns (52.7%). Significant differences in preparedness were observed across geopolitical zones ($\chi^2$ (5) = 24.28, $p<0.001$), and awareness differed across professional groups ($\chi^2$ (6) = 68.38, $p<0.001$). Attitudes toward AI differed significantly across professional groups ($F=3.32$, $p=0.003$), with professionals who felt prepared demonstrating more positive attitudes (mean = 3.74) compared to those who did not (mean = 3.46). These findings reveal a critical disconnect between high awareness and actual readiness, underscoring the need for targeted training, infrastructure investment, and clear implementation frameworks to bridge the gap between AI technological potential and clinical reality in resource-constrained settings.

**Keywords:** Artificial Intelligence; Healthcare Workforce; Readiness; Nigeria; Low-Resource Settings

## 1 Introduction

Artificial intelligence (AI) is increasingly integrated into healthcare systems worldwide, with applications spanning medical imaging, clinical decision support, and predictive analytics [1,2]. Driven by machine learning and deep learning, these technologies have demonstrated substantial potential to improve diagnostic accuracy and streamline clinical workflows [3–5], with algorithms in medical imaging now achieving performance comparable to specialist clinicians [6]. However, successful implementation depends critically on the readiness, willingness, and capacity of healthcare professionals to incorporate AI into routine practice [6,7].

In sub-Saharan Africa, weak digital infrastructure, fragmented health information systems, and governance gaps constrain sustainable AI deployment [8,9]. Nigeria, the continent's most populous nation, faces persistent infrastructural deficits including unreliable electricity, limited rural broadband, and workforce shortages that create a complex environment for AI adoption [10,11]. The concept of workforce readiness is well-theorized: Davis's Technology Acceptance Model (TAM) highlights perceived usefulness and ease of use as primary determinants of acceptance [12], while Venkatesh and colleagues' Unified Theory of Acceptance and Use of Technology (UTAUT) incorporate performance expectancy, effort expectancy, social influence, and facilitating conditions [13]. These frameworks suggest readiness is a multidimensional construct encompassing practical knowledge, institutional support, and hands-on experience [14].

Recent studies across African health systems have documented significant gaps between awareness and actual preparedness. Among African radiographers, while 84.9% believed AI would improve practice, 61.3% expressed fear of job displacement, and 92.5% indicated a need for further education and training [15]. Similarly, a study of Nigerian healthcare workers found that while 85% were familiar with AI concepts, only 54% felt their departments were prepared for integration, and 31% anticipated the need for extensive retraining [16]. These findings underscore a recurring pattern: high awareness coexists with low objective knowledge, limited practical exposure, and significant concerns about workforce implications. Despite growing recognition of these challenges, large-scale, multi-professional assessments of AI readiness remain scarce in the Nigerian context. Existing studies have largely focused on single professional groups primarily radiographers or medical students and have employed limited geographic sampling [15–17]. There is a critical need for comprehensive, nationally representative data that capture the perspectives of diverse healthcare professionals, including physicians, nurses, allied health workers, and technical staff, across Nigeria's varied geopolitical and economic landscapes. Such data are essential for informing policy, designing targeted training interventions, and ensuring that AI deployment strategies are grounded in the realities of frontline clinical practice.

This study addresses this gap by evaluating awareness, attitudes, preparedness, and barriers to AI adoption among a diverse, nationwide sample of Nigerian healthcare professionals. Drawing on established technology acceptance frameworks [12,13], we hypothesized that (1) awareness of AI would be high but objective knowledge and preparedness would be low; (2) attitudes would vary across professional groups and experience levels; and (3) infrastructural and educational barriers would be the most frequently cited obstacles. The findings aim to inform context-specific strategies for workforce development and equitable AI integration in Nigeria and similar LMIC settings.

## 2 Material and Method

### Study design and settings

This was a nationwide cross-sectional survey conducted among healthcare professionals in Nigeria between December 2025 and March 2026. Nigeria is divided into six geopolitical zones (North Central, North East, North West, South East, South South, and South West), each with distinct economic, cultural, and healthcare infrastructure profiles. The study was designed to capture perspectives from all six zones to enable geographic comparisons of AI readiness.

### Participants and Sampling

A total of 761 healthcare professionals participated in the study. Participants were recruited using a combination of convenience and snowball sampling strategies, with targeted outreach to ensure representation across professional disciplines, practice settings (tertiary, secondary, and primary healthcare facilities; public and private sectors), and geopolitical zones. Inclusion criteria were: (1) current employment in a healthcare facility or academic institution in Nigeria; (2) at least one year of professional practice; and (3) provision of informed consent. All professional cadres were eligible, including physicians, nurses, midwives, pharmacists, laboratory scientists, radiographers, physiotherapists, and administrative/technical staff.

### Instrument deployment and validation

The survey instrument was developed following an extensive review of existing literature on AI readiness and technology acceptance in healthcare [12,13,15,16]. The questionnaire comprised five sections: (1) sociodemographic and professional characteristics; (2) awareness and knowledge of AI in healthcare; (3) attitudes toward AI adoption; (4) self-reported preparedness to use AI-based tools; and (5) perceived barriers to AI implementation. Knowledge items were designed to assess both self-perceived knowledge (using Likert scales) and understanding of foundational AI concepts. The objective subscale comprised 7 multiple-choice items (e.g., distinguishing machine learning from rule-based systems), scored 0–1 per item and summed to a 0–7 total. Attitude items were measured on a 5-point Likert scale (1 = Strongly Disagree to 5 =

Strongly Agree) and covered domains including perceived usefulness, trust in AI, concerns about job displacement, and willingness to engage in training. Barrier items were presented as multiple-choice and Likert-scale questions covering infrastructural, financial, educational, ethical, and workforce-related obstacles. Content validity was established through expert review by a panel of five specialists in radiology, medical informatics, health systems research, and biostatistics. The instrument was piloted among 50 healthcare professionals across three states (Kano, Lagos, and Enugu) to assess clarity, comprehension, and internal consistency. Cronbach's alpha coefficients for the attitude, preparedness, and barrier subscales were 0.84, 0.79, and 0.81, respectively, indicating good reliability. Minor revisions were made to wording and item ordering based on pilot feedback.

### Data collection

Data were collected via Google Forms distributed through professional networks, institutional email lists, and social media (WhatsApp, LinkedIn, Twitter/X). Hard copies were distributed at conferences and continuing professional development events to accommodate participants with limited internet access. Weekly reminders were sent. The survey was administered in English only. Data collection spanned 16 weeks.

### Ethical considerations and statistical analysis

Ethical approval was obtained. Participation was voluntary, and all respondents provided electronic or written informed consent. No personally identifiable information was collected; all responses were anonymized prior to analysis.
Statistical analysis was performed in Python 3.12 using Jupyter Notebook, with pandas, SciPy, and statsmodels. Descriptive statistics characterized the sample. Group comparisons used chi-square tests, independent t-tests, and one-way ANOVA with Tukey's HSD post-hoc analyses. Pearson correlations examined relationships among knowledge scores, attitude scores, preparedness ratings, and barrier perceptions. Normality was assessed using the Shapiro-Wilk test. A two-tailed significance level of $p < 0.05$ was used for all inferential tests. A two-tailed $p < 0.05$ was used for all inferential tests. No correction for multiple comparisons was applied across exploratory subgroup analyses.

## 3 Results

### Sociodemographic Characteristics

A total of 761 healthcare professionals completed the survey. The majority of respondents were male (58.2%, n=443), with a mean age of 38.4 ± 9.7 years. Physicians constituted the largest professional group (32.1%, n=244), followed by nurses (24.3%, n=185), radiographers and allied health professionals (18.9%, n=144), laboratory scientists (12.6%, n=96), pharmacists (7.2%, n=55), and administrative/technical staff (4.9%, n=37). Respondents were distributed across all six geopolitical zones, with the highest representation from the North West (24.7%, n=188) and South East (19.6%, n=149). The majority worked in tertiary healthcare facilities (45.6%, n=347), with 31.4% (n=239) in secondary facilities and 23.0% (n=175) in primary care or private practice settings. Professional experience ranged from 1–5 years (28.4%) to more than 20 years (18.1%), with a mean of 12.1 ± 8.4.

## Awareness and Knowledge of AI

Overall awareness of AI in healthcare was 92.6% (n=705). Objective assessment showed 59.1% (n=450) could correctly define basic AI concepts, and 40.9% (n=311) reported their own knowledge as "low" or "very low." Physicians demonstrated significantly higher objective knowledge scores (mean = 4.12 ± 1.34 out of 7) compared to nurses (3.21 ± 1.45), allied health professionals (3.45 ± 1.38), and administrative staff (2.67 ± 1.22; $F(5,755) = 18.47$, $p<0.001$). Awareness differed significantly across professional groups ($\chi^2(6)=68.38$, $p<0.001$). Physicians (98.4%) and radiographers (96.5%) showed the highest awareness, and administrative/technical staff the lowest (78.4%).

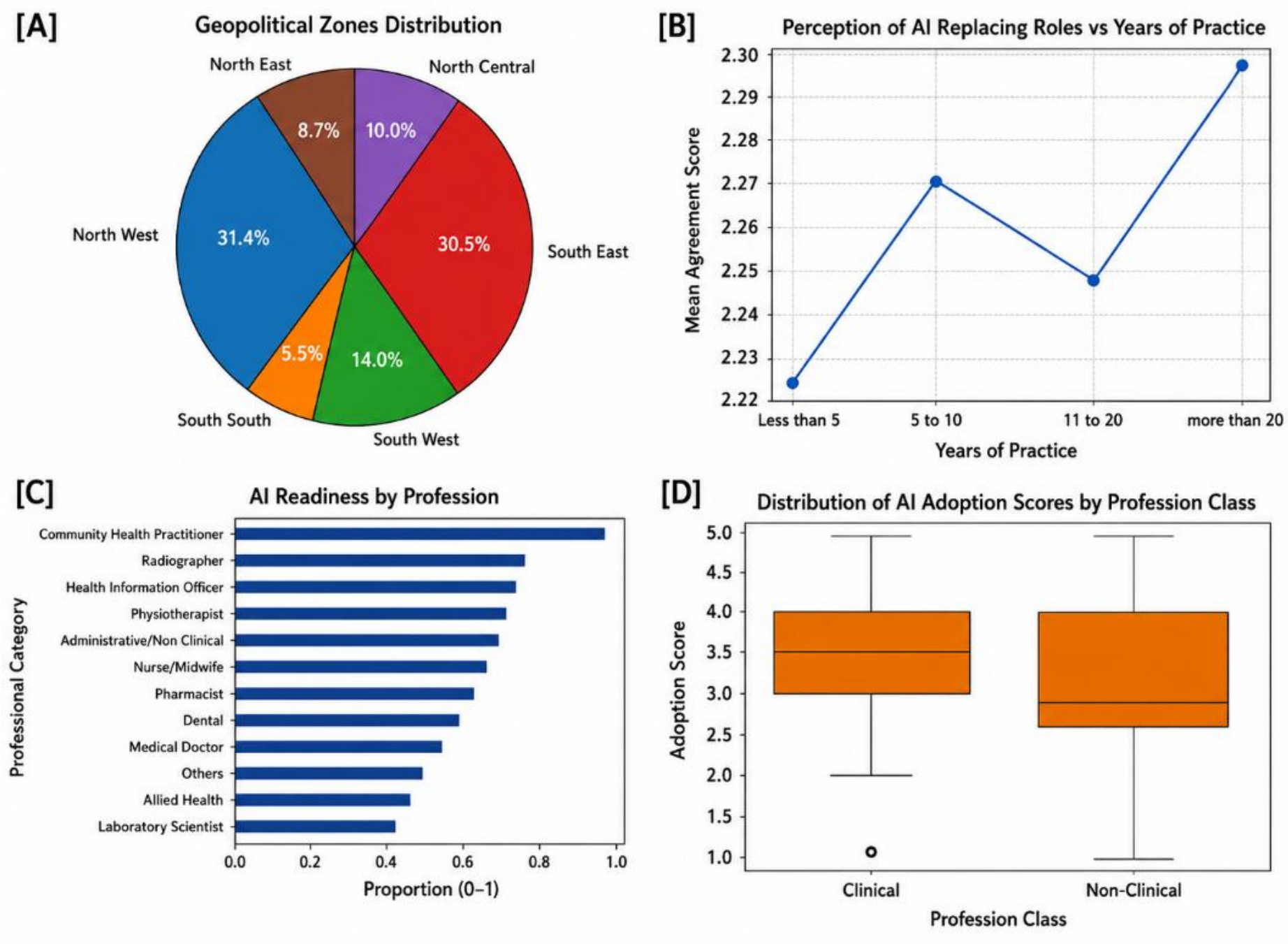

**Fig. 1.** Respondent distribution and perceptions of AI in healthcare. (A) Geopolitical distribution across Nigeria's six zones. (B) Perceived risk of AI replacing healthcare roles by years of experience. (C) AI readiness levels across professional groups. (D) AI adoption scores among clinical and non-clinical professionals.

### Attitudes Toward AI Adoption

Attitudes toward AI varied significantly across professional groups (F=3.32, p=0.003). The overall mean attitude score was 3.58 ± 0.84. Physicians (mean = 3.78) and radiographers (mean = 3.71) reported the highest scores, and administrative staff the lowest (mean = 3.21). Professionals with <10 years' experience expressed a mean score of 3.82, and those with >20 years a mean score of 3.34 (t (759) = 3.89, p<0.001). Professionals who felt prepared demonstrated significantly more positive attitudes (mean = 3.74 ± 0.71) compared to those who did not feel prepared (mean = 3.46 ± 0.91; t (759) = 4.21, p<0.001).

### Preparedness and Willingness

Sixty three percent 63.0% (n=479) of respondents felt adequately prepared. 92.5% (n=704) expressed interest in receiving formal AI training. 82.4% (n=627) indicated willingness to adopt AI for clinical use. 78.7% (n=599) supported inclusion of AI education in undergraduate and postgraduate curricula. Significant differences in preparedness were observed across geopolitical zones ($\chi^2$ (5) =24.28, p<0.001). The South West (71.3%) and South East (68.5%) reported the highest preparedness lev-els, and the North East (52.6%) and North Central (55.1%) the lowest.

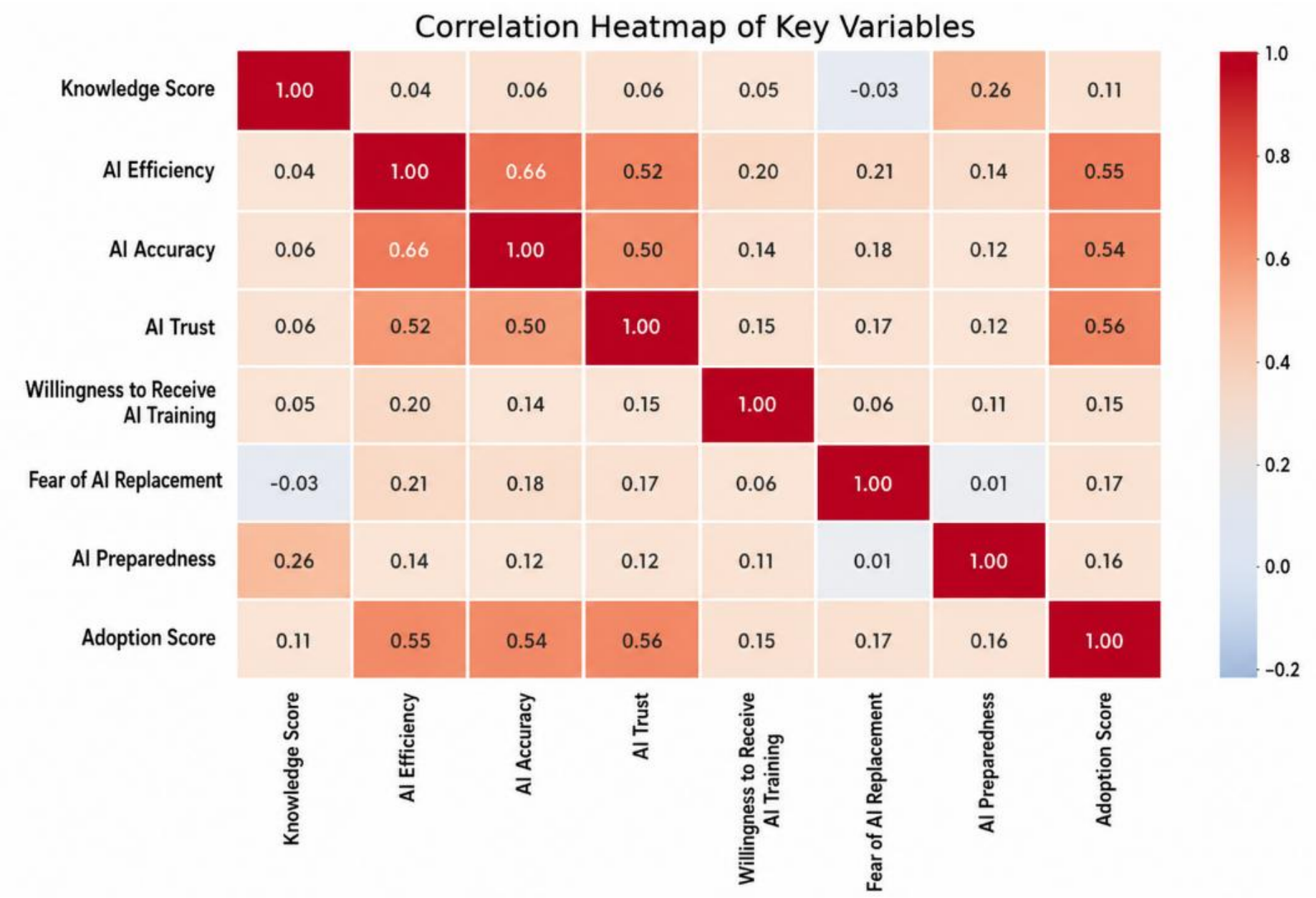

**Fig. 2.** Correlation heatmap of key AI-related variables. Perceived AI efficiency, accuracy, and trust were moderately positively correlated (r = 0.50–0.66). AI adoption scores associated moderately with efficiency (r = 0.55), accuracy (r = 0.54), and trust (r = 0.56). Knowledge scores correlated weakly with other variables, and fear of job displacement showed minimal association with adoption.

### Perceived Barriers to AI Adoption

The most frequently cited barrier was lack of training and education (84.7%, n=645), followed by poor infrastructure (71.1%, n=541), high cost of AI tools (61.0%, n=464), fear of job displacement (60.6%, n=461), ethical concerns (52.9%, n=403), and data privacy concerns (52.7%, n=401). Other barriers included lack of institutional support (48.2%), unreliable internet connectivity (45.6%), and absence of clear regulatory frameworks (41.3%).

**Table 1.** Perceived barriers to AI adoption in clinical practice (n=761). Fear of job displacement increased with years of professional practice. Mean agreement scores were 2.89 ± 1.12 for 1–5 years and 3.67 ± 1.08 for >20 years (F (4,756) =12.43, p<0.001).

| Barrier | n | Percentage (%) | Rank |
|---|---|---|---|
| Lack of training and education | 645 | 84.7 | 1 |
| Poor infrastructure (power, equipment, space) | 541 | 71.1 | 2 |
| High cost of AI tools and implementation | 464 | 61.0 | 3 |
| Fear of job displacement | 461 | 60.6 | 4 |
| Ethical concerns (liability, autonomy) | 403 | 52.9 | 5 |
| Data privacy and security concerns | 401 | 52.7 | 6 |
| Lack of institutional support | 367 | 48.2 | 7 |
| Unreliable internet connectivity | 347 | 45.6 | 8 |
| Absence of clear regulatory frameworks | 314 | 41.3 | 9 |
| Resistance to change among staff | 289 | 38.0 | 10 |

### Intercorrelations Among Key Variables

Moderate positive correlations were observed between perceived AI efficiency, accuracy, and trust (r = 0.50–0.66, p<0.001). AI adoption scores showed moderate positive associations with efficiency (r = 0.55), accuracy (r = 0.54), and trust (r = 0.56). Knowledge scores demonstrated weak correlations with other variables (r = 0.12–0.28).

Fear of AI replacing roles showed minimal association with adoption scores ($r = -0.09$, $p=0.18$) and a moderate negative correlation with trust ($r = -0.34$, $p<0.001$).

## 4 Discussion

Our findings confirm all three hypotheses: awareness of AI is high among Nigerian healthcare professionals, but knowledge and preparedness lag behind; attitudes and readiness vary sharply across professional groups, experience levels, and geopolitical zones; and the dominant barriers are infrastructural and educational, not attitudinal. Respondents know AI is coming many have seen demos, read articles, attended conferences, but knowing a technology exists is not the same as being equipped to use it. This awareness and preparedness gap mirrors patterns across Africa and other LMICs [18]. Botwe and colleagues found African radiographers similarly optimistic yet fearful and undertrained [15], while studies in Ondo State reported high familiarity but low departmental readiness [16]. Our contribution is showing that this paradox generalizes nationwide and across disciplines.

Much of the global conversation around medical AI particularly within the MICCAI community, centers on model performance: accuracy, sensitivity, F1 scores, and benchmark leaderboards. Yet a CT triage algorithm or a chest X-ray screening model that achieves 95% accuracy on a test set is worthless if the radiologist who must use it lacks training to interpret heatmaps, the PACS infrastructure cannot receive AI annotations, or power outages render the workstation inoperable. This study shifts focus from the algorithm to the ecosystem. Our data suggest that Nigeria, like many LMICs, does not primarily suffer from a shortage of clever models; it suffers from a shortage of conditions that allow clever models to be used. The 84.7% who cited lack of training and the 71.1% who cited poor infrastructure are not rejecting AI, they are describing an environment where even the best-intentioned imaging technology will fail to reach patients. For the AFRICAI community, this underscores that workforce readiness and deployment science are not supplementary to model development; they are prerequisites for it. Without parallel investment in clinician-AI interaction design, offline-capable inference, and hospital IT integration, imaging AI innovations risk becoming expensive prototypes that never touch clinical practice.

The UTAUT framework helps explain why enthusiasm does not automatically convert into adoption [13]. Without facilitating conditions reliable power, internet, hardware, institutional protocols, and hands-on training positive attitudes stall. The geographic divide is telling: the North East and North Central zones, already strained by conflict and underinvestment [11], reported the lowest preparedness. If deployment follows the path of least resistance, AI will cluster in better-resourced southern cities and widen existing inequalities. Equity must be engineered into rollout plans from the start.

Four priorities emerge. First, training demand is overwhelming; 92.5% want it. Programs must be practical and workflow-integrated, not theoretical deep-learning courses. Professional bodies should embed AI literacy into continuing education and undergraduate curricula. Second, 71.1% cited poor infrastructure. Offline-capable, edge-deployable models and solar-powered backup systems are not optional luxuries here; they are prerequisites [18,19]. Third, 60.6% fear job displacement, especially senior staff. Change management must be honest AI will reshape work, not eliminate the need for human judgment and senior clinicians should be engaged as co-designers. Fourth, ethical and governance frameworks need strengthening. With 52.9% citing ethical concerns and 52.7% data privacy, national policy must align with WHO and UNESCO guidelines and include enforceable liability and consent standards [20].

Limitations include the cross-sectional design, which precludes causal inference; self-report bias; and English-only administration. Convenience and snowball sampling likely overrepresented AI-aware professionals. The 7-item objective knowledge subscale provides a limited snapshot of AI literacy rather than a comprehensive competency assessment. Given the number of exploratory subgroup comparisons, p-values were not adjusted for multiple testing and should be interpreted descriptively. Future work should use longitudinal designs, intervention studies.

In conclusion, the Nigerian healthcare workforce is willing but not yet ready. The barrier is not resistance; it is the absence of enabling systems. Closing the gap requires coordinated investment in training, resilient infrastructure, and governance. If done equitably, Nigeria can turn high awareness into genuine clinical impact and offer a model for other LMICs facing the same challenge.

**Acknowledgments.** The authors thank the African Institute for Research Advancement & Innovation (AIRA Africa) and the Medical Artificial Intelligence Lab, Lagos, Nigeria (MaiLab.io) for their support of this project.

**Disclosure of Interests.** The authors have no competing interests to declare that are relevant to the content of this article.

## 5 References

1. Severin AM, Radu AC, Rapan I, Poenaru LA. The Role of AI-Driven Technologies in Transforming Healthcare: Trends and Perspectives in Europe and Worldwide. International Journal of Academic Reserach in Economics and Management Sciences. 2025 Aug 31;14(3):479-89.
2. Andigema AS, Cyrielle NN, Ekwelle E. Artificial Intelligence in African Healthcare: Catalyzing Innovation While Confronting Structural Challenges. Preprints. org. 2025.
3. Alqahtani, Tariq & Badreldin, Hisham & Alrashed, Mohammed & Alshaya, Abdulrahman & Alghamdi, Sahar & Saleh, Khalid & Alowais, Shuroug & Alshaya, Omar & Rahman, Ishrat & Al Yami, Majed & Albekairy, Abdulkareem. (2023). The emergent role of artificial intelligence, natural learning processing, and large language models in higher education and research. Research in Social and Administrative Pharmacy. 19. 10.1016/j.sapharm.2023.05.016.
4. Baurasien BK, Alareefi HS, Almutairi Diyanah B, Alanazi Maserah M, Alhasson Aseel H, Alshahrani AD, Almansour SA, Alshagag ZA, Alqattan KM, Alotaibi HM. Medical errors and patient safety: Strategies for reducing errors using artificial intelligence. Int. J. of Health Sci. [Internet]. 2023 Jan. 15 [cited 2026 Jul. 13];7(S1):3471-87. Available from: https://sciencescholar.us/journal/index.php/ijhs/article/view/15143

5. Ahmad S, Wasim S. Prevent medical errors through artificial intelligence: A review. Saudi J Med Pharm Sci. 2023 Jul 11;9(7):419-23.
6. Alowais, S.A., Alghamdi, S.S., Alsuhebany, N. *et al.* Revolutionizing healthcare: the role of artificial intelligence in clinical practice. *BMC Med Educ* **23**, 689 (2023). https://doi.org/10.1186/s12909-023-04698-z
7. Mollura DJ, Culp MP, Pollack E, Battino G, Scheel JR, Mango VL, Elahi A, Schweitzer A, Dako F. Artificial Intelligence in Low- and Middle-Income Countries: Innovating Global Health Radiology. Radiology. 2020 Dec;297(3):513-520. doi: 10.1148/radiol.2020201434. Epub 2020 Oct 6. PMID: 33021895.
8. Oleribe OO, Momoh J, Uzochukwu BS, Mbofana F, Adebiyi A, Barbera T, Williams R, Taylor-Robinson SD. Identifying Key Challenges Facing Healthcare Systems In Africa And Potential Solutions. Int J Gen Med. 2019 Nov 6;12:395-403. doi: 10.2147/IJGM.S223882. PMID: 31819592; PMCID: PMC6844097.
9. International Centre of Expertise in Montreal on Artificial Intelligence (CEIMIA). (2024).State of AI in Healthcare in Sub-Saharan Africa.https://doi.org/10.5281/zenodo.12628185
10. Alami, H., Rivard, L., Lehoux, P. *et al.* Artificial intelligence in health care: laying the Foundation for Responsible, sustainable, and inclusive innovation in low- and middle-income countries. *Global Health* **16**, 52 (2020). https://doi.org/10.1186/s12992-020-00584-1
11. Antwi WK, Akudjedu TN, Botwe BO. Artificial intelligence in medical imaging practice in Africa: a qualitative content analysis study of radiographers' perspectives. Insights Imaging. 2021 Jun 16;12(1):80. doi: 10.1186/s13244-021-01028-z. PMID: 34132910; PMCID: PMC8206887.
12. Davis FD. Perceived usefulness, perceived ease of use, and user acceptance of information technology. MIS quarterly. 1989 Sep 1;13(3):319-40.
13. Venkatesh, Viswanath & Morris, Michael & Davis, Gordon & Davis, Fred. (2003). User Acceptance of Information Technology: Toward A Unified View1. MIS Quarterly. 27. 425-478. 10.2307/30036540.
14. Akudjedu TN, Torre S, Khine R, Katsifarakis D, Newman D, Malamateniou C. Knowledge, perceptions, and expectations of Artificial intelligence in radiography practice: A global radiography workforce survey. Journal of Medical Imaging and Radiation Sciences. 2023 Mar 1;54(1):104-16.
15. Al-Ganad A, Al-Shahdhi A, Al-Dhaifi O, Hajeb E, Hajeb H and Al-Motarreb A (2026) Deploying medical AI in low-resource settings: a scoping review of challenges and strategies. Front. Digit. Health 8:1743634. doi: 10.3389/fdgth.2026.17436344.
16. Alatawi A, Ali D. AI Implementation in Healthcare Industry: A Systematic Review. JRR [Internet]. 2025 Dec. 19 [cited 2026 Jul. 13];2:57-71. Available from: https://journalrr-site.com/index.php/Myjrr/article/view/191
17. Abuzaid MM, Elshami W, McConnell J, Tekin HO. An extensive survey of radiographers from the Middle East and India on artificial intelligence integration in radiology practice. Health Technol (Berl). 2021;11(5):1045-1050. doi: 10.1007/s12553-021-00583-1. Epub 2021 Aug 6. PMID: 34377625; PMCID: PMC8342654.
18. World Health Organization: Ethics and Governance of Artificial Intelligence for Health: WHO Guidance. WHO, Geneva (2021)
19. UNESCO: Recommendation on the Ethics of Artificial Intelligence. UNESCO, Paris (2021)
20. Zaidan AM. The leading global health challenges in the artificial intelligence era. Front Public Health. 2023 Nov 27;11:1328918. doi: 10.3389/fpubh.2023.1328918. PMID: 38089037; PMCID: PMC10711066.